\documentclass{sig-alternate}
\begin{document}
%%
%% --- Author Metadata here ---
%\conferenceinfo{RACS'13}{October 1-4, 2013, Montreal, QC, Canada.}
%\CopyrightYear{2013} % Allows default copyright year (20XX) to be over-ridden - IF NEED BE.
%\crdata{0-12345-67-8/13/10}  % Allows default copyright data (0-89791-88-6/97/05) to be over-ridden - IF NEED BE.
%% --- End of Author Metadata ---

\title{Energy Aware Error Control in Cooperative Communication in Wireless Sensor Networks}

\author{B. Manzoor$^{\ddag}$, N. Javaid$^{\ddag, \$}$, O.Rehman$^{\ddag}$, S. H. Bouk$^{\ddag}$, S. H. Ahmed$^{\sharp}$, D. Kim$^{\sharp}$\\\\
$^{\ddag}$EE Department, COMSATS Institute of Information Technology, Islamabad, Pakistan. \\
        $^{\$}$CAST, COMSATS Institute of Information Technology, Islamabad, Pakistan.\\
        $^{\sharp}$School of Computer Science \& Engineering, Kyungpook National University, Korea.\\
}

\maketitle
\begin{abstract}
Due to small size of sensor nodes deployed in Wireless Sensor Networks (WSNs), energy utilization is a key issue. Poor channel conditions lead to retransmissions and hence, result in energy wastage. Error control strategies are usually utilized to accommodate channel impairments like noise and fading in order to optimize energy consumption for network lifetime enhancement. Meanwhile, cooperative communication also emerges to be an appropriate candidate to combat the effects of channel fading. Energy efficiency of cooperative scheme when applied with Automatic Repeat Request (ARQ), Hybrid-ARQ (HARQ) and Forward Error Correction (FEC) is investigated in this work. Moreover, the expressions for energy efficiency of Direct Transmission, Single Relay Cooperation and Multi Relay Cooperation are also derived. In all, our work is focused towards energy optimal communication in WSNs. Our results show that error control strategies with cooperative schemes can significantly enhance system performance in form of energy optimization.
\end{abstract}


\section{Introduction}
\label{sec:intro}

Technological advancements opened many platforms for researchers to pave paths for miniaturization. WSNs due to their importance for many applications like monitoring, signal processing, health and etc., gained a tremendous attention in the recent past. Sensor nodes deployed in WSNs are miniature in size and are capable of efficient sensing, processing and communication. Due to their low size they are power constrained as life-time of battery strictly depends upon the size of device. Therefore, main goal in WSNs is to enhance network lifetime by efficient utilization of available energy resources~\cite{10}.

There are many strategies devised for the optimization of energy consumption in WSNs. Switching transceiver to \textit{on} and \textit{off} mode is considered one of the way to improve energy efficiency, as transceiver is a most energy consuming module in a communication system. Another way to reduce energy consumption is to improve network routing algorithms for load balancing among sensor nodes in a network. However, the above mentioned solutions are prone to channel impairments and hence, result in degradation of system performance ~\cite{2, 4, 13, 15} %~\cite{pellenz2010error}.

Error control coding (ECC) is a well defined approach that lowers the required transmit power. Strong codes provide better performance at the cost of complex circuitry on the decoder sides which is a wastage of battery resources ~\cite{1}. Depending upon application requirements, performance of FEC codes and ARQ is addressed in ~\cite{8}. Hybrid ARQ which is combination of conventional ARQ with FEC is well investigated in ~\cite{3,9}. A soft decision relaying function Estimate-Forward (EF) is applied for relaying networks. Based upon mutual information (MI) throughput efficiency of HARQ is investigated in this work. Moreover, in ~\cite{7} a detailed study is carried to investigate the packet transfer delay and energy efficiency of two basic retransmissions, which are hop-by-hop and end-to-end retransmissions.

Recent research shows that Cooperative Diversity emerges to be a potential candidate to combat the effects of channel fading via cooperation among relay nodes. Optimal incremental relaying (IR) cooperation strategy is investigated in ~\cite{6} and energy efficiency of cooperative diversity for WSNs is discussed in detail.

In this paper, we tried to investigate energy efficiency of ARQ, HARQ and FEC techniques for WSNs. Main focus of our work is to show impact of relay communication when implemented along with these schemes. Compared with direct transmission (DT) and single relay cooperation (SRC) in this paper, multi relay cooperation (MRC) proves in optimizing transmission energy for successful data transmission. Successful data transmission in SRC involves three transmitting nodes, the source, and two relay nodes. However, channel conditions between source and destination and source and relays along with distance separation between them are also important factors to be considered. We tried to evaluate performance on the basis of different parameters like bit error rate (BER), symbol error rate (SER), Delay and Throughput.

The remainder of this paper is organized as follows. Section II presents an overview of previous work on error correction and detection along with importance of cooperative communication applied to WSNs. In section III ECCs are discussed and their energy consumption is highlighted. System model with expressions of energy efficiency for DT, SRC and MRC are discussed in section IV of this paper. Section V gives a detailed discussion of simulations result and finally in section VI we concluded our paper.

\section{Related Work and Motivation}
\label{sec:format}

A lot of work is carried which focuses on energy efficiency optimization problem in WSNs (e.g.~\cite{11, 12, 14, 16}). Different types of ECCs are investigated in ~\cite{2} and based upon BER and power consumption criteria, Reed Solomon code $RS(31,21)$ is considered to be an appropriate choice. Similarly, trade-off between transmission and processing energy consumption is investigated in ~\cite{4}. In this study, an hybrid scheme based on ARQ and FEC is studied and link error rates between different pair of network nodes are considered as an evaluation metric. FEC schemes employing convolutional codes with code rate $1/2$ are considered for evaluation.

Energy expenditure per data bit for both coded and un-coded system is discussed in ~\cite{1}. This energy expenditure is taken as a function of critical distance $d_{CR}$ and it is concluded that at distances greater than $d_{CR}$, use of coded system results in net energy savings for WSNs. A cross-layer analysis of error control in wireless sensor networks is addressed in ~\cite{8}. A comprehensive comparison of FEC and ARQ is studied and cross layer effect of routing, medium access and physical layer is considered.

An incremental redundancy IR-HARQ protocol is discussed in detail by ~\cite{3}. Possible use of Low Density Parity Check (LDPC) and Raptor codes for HARQ is studied in detail in this work. In order to avoid retransmission for efficient usage of resources through memoryless relaying is defined by ~\cite{9}. In this study, non-decoding relays are introduced for resource management. Similarly number of retransmissions resulting from frame error are identified in ~\cite{5}. Moreover, this study signifies applicability of advance FEC schemes in WSNs.

\begin{figure*}[!ht]
\centering
%{\includegraphics[scale=0.5]{FEC.eps}}
 \includegraphics[height=8cm,width=14 cm]{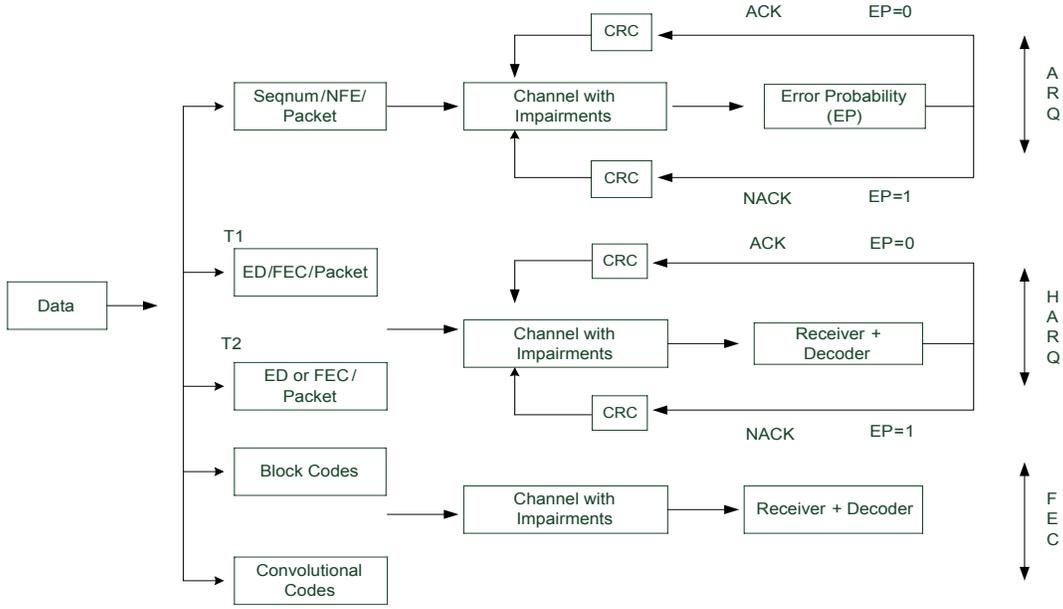}
 \vspace{-.4cm}
\caption{Description of Coding Schemes}
\end{figure*}

Fig. 1 represents working principle of ARQ, HARQ and FEC.  From figure it can be seen that data is transmitted in the form of frames by transmitter. Depending upon error probability (EP) receiver decides to reply with acknowledgement (ACK) for error free frame transmission or with negative-ACK (NACK) for retransmission. As ACK and NACK uses same channel, therefore,  they must also be protected by cyclic redundancy check (CRC). In case of lost or delayed frame sender must time out and resend the previous frame. Moreover, sequence number (SeqNum) is used to avoid the reception of duplicate frames. SeqNum and next frame expected (NFE) values are added in the frame header.

In type 1 (T1) HARQ  error detection (ED) information along with FEC are added before transmission. On reception of coded data block receiver decodes error correction code. In case of good channel quality transmission errors should be correctable, and correct data block is received. Through error correction codes channel condition are examined and in case of bad condition a retransmission is requested following the ARQ. In type 2 (T2) HARQ on successive transmissions, either ED bits or FEC information bits are transmitted.

To avoid retransmissions FEC is used in data transmission. Through addition of redundant data into its message allows receiver to detect and correct errors within certain bounds. FEC is categorized into two categories Block coding and Convolutional coding. Block codes works on the fixed size blocks of data  whereas, convolutional codes work on data streams of arbitrary lengths. In block codes information bits are followed by parity bits whereas, in case of convolutional codes information bits are spread along the sequence.
\begin{figure}[!ht]
  \centering
   \includegraphics[height=4cm,width=7 cm]{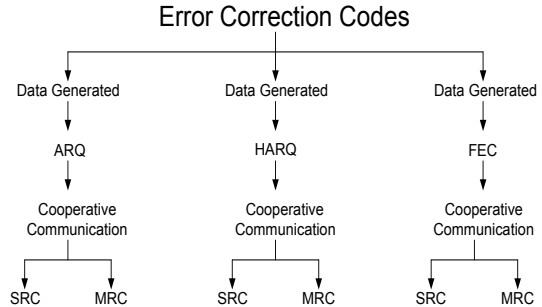}
    \vspace{-.2cm}
  \caption{Hierarchy of Coding Schemes with SRC and MRC}
\end{figure}

Fig. 2 illustrate an example of error correction and detection schemes. Figure shows that ARQ, HARQ and FEC can be used as ECC for better performance of communication system. Depending upon application, any of the these schemes can be implemented for efficient and reliable communication. Effect of cooperative communication in these schemes is also shown through figure. SRC and MRC can also be implemented in order to utilize energy of sensor nodes more efficiently. Original data is encoded with any of the coding schemes then it is transmitted following the principle of cooperative communication and on the reception side it is decoded accordingly.

Motivated by previous work for energy efficiency of WSNs, impact of cooperative communication when applied with ARQ, HARQ and FEC is investigated in this paper. Expressions for energy efficiency of SRC and MRC are also derived. Moreover, performance of above mentioned schemes is studied in detail. We summaries our contribution in a way that use of MRC can significantly enhance system output. Through MRC, encoded data to be transmitted is divided into sub-frames and then transmitted through multiple paths. This multi-path transmission of data not only reduces cost of retransmissions but also adds in energy efficiency of the system. The results show that our proposed scheme can play a vital role in WSNs.

\section{Error Detection and Correction}

Redundancy introduced into an information sequence through addition of extra parity bits allows the decoder to possibly decode noisy received bits correctly. The ability to correct errors in the received sequence is normally defined as ECC. Through the usage of ECC better BER performance for the same signal-to-noise (SNR) ratio when compared with un-coded system can be achieved, or can provide the same BER at a lower SNR than un-coded. SNR difference to achieve a certain BER for a particular code and decoding algorithm compared to un-coded is defined as coding gain for that code and decoding algorithm.

Minimum transmit power required for an un-coded system to achieve a desired BER at a given SNR can be found by ~\cite{1} as:

\begin{eqnarray}\label{eq:1}
P_{TX,U}[W]=\eta u10^{(SNR_u/10+RNF/10)}(KTB)\frac{E_b}{N_o}N{(\frac{4\pi}{\lambda})}^2 d^n
\end{eqnarray}

where, RNF defines receiver noise in dB, $\eta u$ is un-coded system's spectral efficiency and $SNR_U$ defines required SNR to achieve the target BER with an un-coded system. Required minimum transmit power when using ECC is given by:

\begin{eqnarray}
P_{TX,ECC}[W]=\frac{\eta_c B_c}{\eta u B}\frac{P_{TX},U}{10^{ECC_{gain}/10}} = \frac{P_{TX},U}{10^{ECC_{gain}/10}}
\end{eqnarray}

where, $Eb_{TX} = P_{TX}/R$ in J/bits is the required transmit energy per transmitted information bit, and is obtained by dividing required transmit power by $P_{TX}$ with transmission R in bps. Difference between the minimum required transmit energy per information bit for coded and un-coded system defines transmit energy saving per information bit of the coded system and is given as:

\small
 \begin{eqnarray}
 \begin{split}
Eb_{TX,U}[j/bit] &=\frac{P_{TX,U}}{R} \\
Eb_{TX,ECC}[j/bit] &=\frac{P_{TX,ECC}}{R} = \frac{Eb_{TX,U}}{10^{ECC_{gain}/10}} \\
Eb_{TX,U}-Eb_{TX,ECC} &= Eb_{TX,U}(1-10^{-ECC_{gain}/10})
\end{split}
\end{eqnarray}
\normalsize

Therefore, use of ECC optimize the required minimum transmit power and energy per decoded bit as a result of the coding gain $ECC_{gain}$.

\section{System Model}
In this section, we discuss energy efficiency of DT, SRC and MRC. Moreover, expressions for energy consumption are also derived for all of these mentioned strategies.

\begin{figure}[!ht]
\centering
{\includegraphics[scale=0.25]{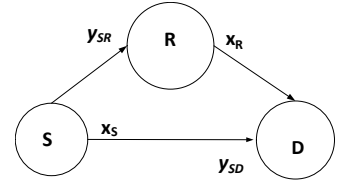}}
 \vspace{-.4cm}
\caption{System model for SRC}
\end{figure}

We consider two different cases of relay communication named as SRC and MRC for WSNs. Meanwhile, we also tried to evaluate the performance of ARQ HARQ and FEC when applied with these two different scenarios.

Fig. 3 shows system model for SRC. In terms of energy efficiency, DT, SRC and MRC are analyzed. Considering all possible sources of energy consumption, let $\beta$ be the loss factor of power amplifiers ranging from $0<\beta<1$ and $P_{ct}$ and $P_{cr}$ are the power consumptions of transmitting and receiving circuitry. Assuming symbol rate $R_s$ to be constant, bit rate will be equal to $R_{s} \ast b$ .

\begin{figure}[!ht]
\centering
{\includegraphics[scale=0.25]{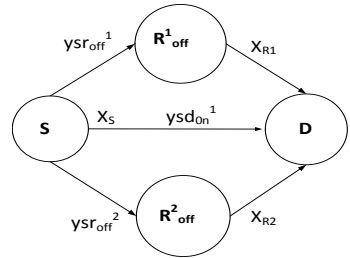}}
 \vspace{-.4cm}
\caption{System model for MRC}
\end{figure}

Fig. 4 with source \textit{S}, destination \textit{D} and off-path relays $R_{off}$ shows system model for MRC. The transmitted signal in the form of symbols will be denoted as $x_s$. Due to broadcast nature of wireless medium, $R_{off}$ also overhears this communication. The received signal at direct and $R_{off}$ paths will be:

\begin{eqnarray}
y_{SD}=h_{SD}x_{S} + n_{SD}
\end{eqnarray}

\begin{eqnarray}
y_{SRoff}^1=h_{SRoff}^1x_{S} + n_{Roff}^1
\end{eqnarray}

\begin{eqnarray}
y_{SRoff}^2=h_{SRoff}^2x_{S} + n_{Roff}^2
\end{eqnarray}

where, $y_{SRoff}$ and $y_{SD}$ denote channel coefficients, $n_{Roff}$ and $n_{Ron}$ denote AWGN at the input of off-path and on-path relays. Therefore, received signal at the destination originated from source will be equal to:

\begin{eqnarray}
y_{RD}=y_{SD}+ y_{SRoff}^{1}+y_{SRoff}^{2}
\end{eqnarray}

Assuming that transmitting power for all nodes is constant and is denoted by $P_t$, $\alpha$ is the path loss exponent and noise components being modeled as additional AWGN with variance $N_o$. Received SNR $\gamma$ for a link can be obtained by probability distribution function (PDF) given in ~\cite{6} as:

\begin{eqnarray}
f_{\gamma ij}(\gamma)= \frac{1}{\sigma_{ij}} exp (-\frac{\gamma}{\sigma_{ij}})     ((i,j)=(sd),(rd),(sr))
\end{eqnarray}

Average SNR $\sigma_{ij}$ and can be expressed as:
\begin{eqnarray}
\sigma_{ij}=\frac{P_{t}(\gamma_{ij})^{-\alpha}}{N_0}
\end{eqnarray}

where, (ij) denotes the different links and $\gamma_{ij}$ denotes distance of a link with node $i$ and $j$ being transmitter and receiver, respectively.
Assume that $M-QAM$ is adapted with the modulation level $b=log_2M bits/symbol$, the closed-form expression for the average symbol error-rate (SER) of a link is given by ~\cite{6} as:
 \begin{eqnarray}
SER_{ij}\approx2(1-2^{-b/2})(1-\sqrt{\frac{3\sigma_{ij}}{2(2^b-1)+3\sigma_{ij}}})
\end{eqnarray}

Thus, PER of a link can be obtained as:
 \begin{eqnarray}
PER_{ij}=1-(1-SER_{ij})^{L/b}
\end{eqnarray}

where, L is the length of the data packet. Considering, PER of DT equal to PER of S-D link it can be written as:
  \begin{eqnarray}
PER^D=PER_{sd}=1-(1-SER_{sd})^{L/b}
\end{eqnarray}

PER for SRC given by ~\cite{6} can be evaluated as:
  \small
  \begin{eqnarray}
PER_{SRC}=PER_{sd}PER_{sr}+PER_{sd}(1-PER_{sr})PER_{rd}
\end{eqnarray}
\normalsize

Based upon (13), we calculate PER for MRC as follows:
\small
 \begin{eqnarray}
 \begin{split}
PER_{MRC} &=PER_{sd}PER_{sr1}PER_{sr2} \\
&+PER_{sd}(1-PER_{sr})PER_{r1d}PER_{r2d}
\end{split}
\end{eqnarray}
\normalsize

From above equation it is clear that transmission from S to D can be carried through multiple ways i.e., multiple paths like S-D and S-R-D are used for efficient communication and corresponding PER can be reduced. Moreover, energy efficiency of the overall system is expressed as:
 \begin{eqnarray}
\eta= \frac {L_p(1-PER)}{E}
\end{eqnarray}

where, $L_p$ is the data packet length, PER denotes packet error rate of DT, SRC and MRC and E is the energy consumption for transporting a data packet with either of the above mentioned schemes.
Now, total energy, consumed for transmitting one data packet with DT is expressed as:
 \begin{eqnarray}
E^D=(P_t(1+\beta)+P_{ct}+p_{cr})\frac{L}{R_b}
\end{eqnarray}

Energy efficiency of DT is given by:
 \begin{eqnarray}
\eta^D= \frac {L_p(1-PER^D)}{E^D}
\end{eqnarray}

Total consumed power for SRC to transmit a single packet is statistically described by ~\cite{6} as:
\small
\begin{eqnarray}
 P^{SRC}_{total}=\left\{\begin{matrix}
 p_{t}(1+\beta)+P_{ct}+2P_{cr}&(1-PER_{sd}), \\
 P_t(1+\beta)+P_{ct}+2P_{cr}&(PER_{sd}PER_{sr}), \\
 2P_t(1+\beta)+2P_{ct}+3P_{cr}&(PER_{sd}(1-PER_{sr})). \\
\end{matrix}\right.
\end{eqnarray}
\normalsize

The first term in above expression $(1-PER_{sd})$ defines successful transmission over S-D link and  receiving power of D and R is denoted by $2P_{cr}$. In the same way $(PER_{sd}PER_{sr})$ defines failure probability of both transmissions over the S-D and S-R links. At the end $(PER_{sd}(1-PER_{sr}))$ defines failure of transmission over an S-D link.
Therefore, the total consumed energy of transmitting one data packet with SRC is given as:

\begin{eqnarray}
\begin{split}
E^c &= \frac{(1-PER_{sd})(P_t(1+\beta)+P_{ct}+2P_{cr})L}{R_b} \\
&+\frac{PER_{sd}PER_{sr}(P_t(1+\beta)+P_{ct}+2P_{cr})L}{R_b} \\
&+\frac{PER_{sd}(1-PER_{sr})(2P_t(1+\beta)+2P_{ct}+3P_{cr})L}{R_b}
\end{split}
\end{eqnarray}
\normalsize

Thus, energy efficiency of SRC is given by:
\begin{eqnarray}
\eta^{SRC}=\frac{L_p(1-PER^{SRC})}{E^{SRC}}
\end{eqnarray}
Therefore, using (18) we calculate total consumed power for MRC in order to transmit one data packet as follows:
\tiny
\begin{eqnarray}
P^{MRC}_{total}=\left\{\begin{matrix}
 p_{t}(1+\beta)+P_{ct}+3P_{cr}&(1-PER_{sd}), \\
 P_t(1+\beta)+P_{ct}+3P_{cr}&(PER_{sd}PER_{sr1}PER_{sr2}), \\
 3P_t(1+\beta)+2P_{ct}+3P_{cr}&(PER_{sd}(1-PER_{sr1}PER_{sr2})). \\
\end{matrix}\right.
\end{eqnarray}
\normalsize

In above expression the term $3P_{cr}$ defines receiving power of D, $R^1_{off}$ and $R^2_{off}$ and $3P_t$ defines transmit power of S, $R^1_{off}$ and $R^2_{off}$, respectively. Moreover, through (19) total consumed energy for transmitting one data packet with MRC is calculated as:
\small
\begin{eqnarray}
\begin{split}
E^c &= \frac{(1-PER_{sd})(P_t(1+\beta)+P_{ct}+3P_{cr})L}{R_b} \\
&+\frac{PER_{sd}PER_{sr1}PER_{sr2}(P_t(1+\beta)+P_{ct}+3P_{cr})L}{R_b} \\
&+\frac{PER_{sd}(1-PER_{sr1}PER_{sr2})(2P_t(1+\beta)+3P_{ct}+3P_{cr})L}{R_b}
\end{split}
\end{eqnarray}
\normalsize

Thus, energy efficiency of MRC is calculated as:
\begin{eqnarray}
\eta^{MRC}=\frac{L_p(1-PER^{MRC})}{E^{MRC}}
\end{eqnarray}

To understand the importance of relay communication in WSNs for enhancement of network life we applied these schemes along with ARQ, HARQ and FEC.
We considered the above system model and tried to investigate the effect of relay communication in three different techniques named as ARQ, HARQ and FEC.

\section{Performance Evaluation}
In this section, we present simulations result of ARQ, HARQ and FEC schemes. Moreover, we also present the effect of off-path relays in the form of SRC and MRC for resource optimization of WSNs. In our simulations specialized form of Reed Solomon (RS) code is used as FEC. Our coding scheme assumes a code of RS(N,K) resulting in a codeword of length N symbols and each having K bits of data in it. Stop and Wait ARQ with CRC-4 is also analyzed. Additionally we also investigate the impact of SRC and MRC on HARQ. The $(7,4)$ Hamming code is used in HARQ. We used MATLAB for our simulations.

\begin{figure}[!ht]
  \centering
  \includegraphics[scale=0.45]{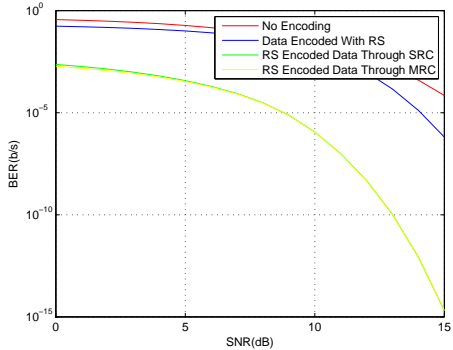}
   \vspace{-.5cm}
  \caption{Performance evaluation of REED SOLOMON codes using SRC and MRC}
\end{figure}

Fig. 5 gives detailed description of RS codes being used as FEC. We try to demonstrate effect of FEC encoding along with cooperative communication. Hard decision RS encoder is used as FEC. BER analysis is carried in the presence of AWGN channel. From figure it is clear that data transmitted without any sort of coding has higher BER. Data when transmitted with RS code encryption achieves better BER. Figure also illustrates impact of SRC and MRC along with FEC encoding. It is seen that use of SRC and MRC significantly enhances system output and achieves much better BERs when compared with un-coded data and the data sent without cooperation. Introduction of SRC and MRC in a communication system can yield in an optimized performance.
\begin{figure}[!t]
  \centering
  \includegraphics[scale=0.45]{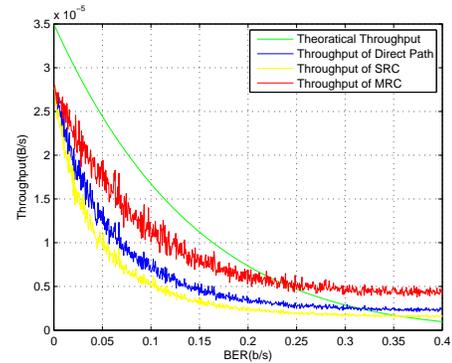}
  \vspace{-.5cm}
  \caption{Throughput analysis of ARQ using SRC and MRC}
\end{figure}

In Fig. 6 throughput analysis is shown as a function of BER for ARQ scheme. Throughput for Stop and wait ARQ along with DT, SRC and MRC is calculated. From figure it is clear that at lower bit error rates throughput of direct path and SRC degrades whereas, MRC follows certain trends. MRC seems to be a good choice in both lower and higher BERs. Furthermore, comparison of these approaches with theoretical throughput also signifies the advantage of MRC more explicitly.
\begin{figure}[!ht]
  \centering
  \includegraphics[scale=0.45]{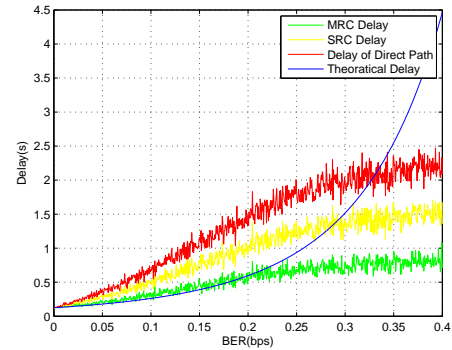}
   \vspace{-.5cm}
  \caption{Analysis of Delay in ARQ using SRC and MRC}
\end{figure}
%\vspace{-2cm}
Fig. 7 shows analysis of ARQ scheme. Stop and wait ARQ with CRC-4 detection is used in our simulations. Figure depicts system BER versus transmission Delay. Delay of DT, SRC, and MRC is calculated and compared against theoretical results. From figure it is clear that in case of off-path relay, transmission delay may vary substantially depending upon relay position. Whereas in case of direct communication this variation can be avoided. Moreover, use of MRC can enhance system performance remarkably.

\begin{figure}[!ht]
  \centering
   \includegraphics[scale=0.5]{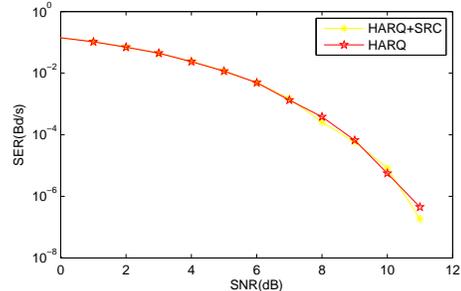}
     \vspace{-.5cm}
  \caption{Performance analysis of HARQ with SRC}
\end{figure}

Fig. 8 shows SER analysis of HARQ scheme using SRC. From figure it is clear that for a given value of SNR a limited improvement is observed  in SER. This happens due to errors caused by fading over a single channel.
\begin{figure}[!ht]
  \centering
  \includegraphics[scale=0.5]{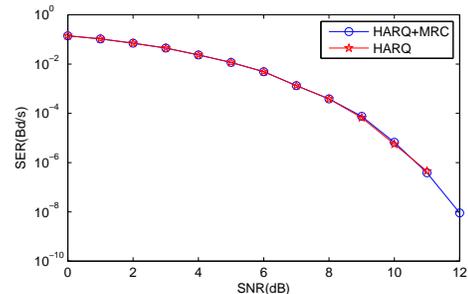}
   \vspace{-.5cm}
   \caption{Performance analysis of HARQ with MRC}
\end{figure}
Fig. 9, shows analysis of HARQ scheme implemented using MRC. We consider source and relay has equal power level for transmission. $T2$ HARQ is investigated in our simulations. From figure it is clear that better symbol error rates are achieved when visualized in a given set of operating SNR.

\section{Conclusion and Future Work}
\label{sec:majhead}

In this work, energy efficiency of three different error correction and detection strategies when implemented with cooperative diversity is studied and compared. We investigated the use of ARQ, HARQ and FEC for error correction and detection in WSNs. Moreover, we also investigated significance of SRC and MRC when applied along with these schemes. The results reveal that ARQ, HARQ and FEC along with SRC and MRC has significantly enhanced system performance. We evaluated the performance on the basis of Throughput, Delay, BER and SER. Network topology is an important feature in energy optimization of WSNs, i.e., optimal selection of relays for efficient communication is also to be taken in account. Therefore, in future we will extend our work in such way that it can support denser networks more efficiently. In future we are interested to implemented the error correction codes in \cite{17-20}.

%\bibliographystyle{abbrv}
%\bibliography{mybibliography}

\end{document}